\documentclass[prd,nofootinbib]{revtex4}
\usepackage{amssymb}
\usepackage{natbib}

\usepackage[T1]{fontenc}

\def\vec#1{\ensuremath{\mathchoice{\mbox{\boldmath$\displaystyle#1$}}
{\mbox{\boldmath$\textstyle#1$}}
{\mbox{\boldmath$\scriptstyle#1$}}
{\mbox{\boldmath$\scriptscriptstyle#1$}}}}

\def\be{\begin{equation}}
\def\fe{\end{equation}}
\def\bea{\begin{eqnarray}}
\def\fea{\end{eqnarray}}

\def\nix{{\rule{0em}{1em}}}

\def\O{{\mathcal{O}}}
\def\text#1{\textrm{#1}}
\def\mb{\mbox}

\def\X{{\mathit{X}}}
\def\Y{{\mathit{Y}}}

\def\Lagr{{\Lambda}}
\def\G{\mathrm{grav}}

\def\EM{{\scriptscriptstyle{\mathrm{EM}}}}
\def\H{{\scriptscriptstyle{\mathrm{H}}}}

\def\Jem{{J_\EM}}

\def\LagrH{{\Lagr_\H}}
\def\LagrF{{\Lagr_\EM}}
\def\LagrG{{\Lagr_\G}}

\def\Act{{\mathcal{I}}}

\def\d{\delta}
\def\D{\Delta}
\def\qaq{{\quad\text{and}\quad}}
\def\E{{\mathcal{E}}}

\def\Dv{\Delta}
\def\entr{\alpha}
\def\entrn{\varepsilon}

\def\vx{{\vec{x}}}
\def\vxi{ {\vec{\xi}} }

\def\dtau{{\tau}}
\def\dtauA{{\dtau_\X}}

\def\e{{\rm e}}

\def\eps{\epsilon}
\def\ext{{\mathrm{ext}}}

\def\vn{{\vec{n}}}
\def\vv{{\vec{v}}}
\def\vnabla{{{\nabla}}}
\def\vDv{{\vec{\Dv}}}
\def\vp{\vec{p}}

\def\vpi{{\vec{\pi}}}

\def\Rate{\mathit{\Gamma}}
\def\csum{\sum}
\def\Press{\Psi}

\def\S{{\mathrm{S}}}

\def\cf{f}
\def\vcf{\vec{\cf}}

\def\vf{\vec{f}}
\def\vfH{\vf_\H}
\def\vfF{\vf_\EM}
\def\vfG{\vf_\G}

\def\FF{{\mathfrak{F}}}

\def\vV{{\vec{V}}}
\def\vb{{\vec{\beta}}}

\def\vort{w}
\def\Vort{W}
\def\cvort{\varpi}
\def\cVort{{\mathcal{W}}}

\def\Circ{{\mathcal{C}}}

\def\Lie{\pounds}
\def\form{\underline}

\def\Hel{{\mathcal{H}}}

\def\Ao{{A_0}}
\def\vA{{\vec{A}}}
\def\vj{{\vec{j}}}

\def\vJ{{\vec{J}}}
\def\vQ{{\vec{Q}}}
\def\vS{{\vec{S}}}

\def\vE{{\vec{E}}}
\def\vB{{\vec{B}}}
\def\vD{{\vec{D}}}
\def\vH{{\vec{H}}}

\def\jo{{\sigma}}

\def\1{{(1)}}
\def\2{{(2)}}

\def\conduct{{\mathfrak{c}}}
\def\ee{\varepsilon}

\def\R{{\!\!R}}

\begin{document}

\title{Variational description of multi-fluid hydrodynamics: Coupling to gauge fields} 

\author{Reinhard Prix}

\affiliation{Max-Planck-Institut f\"ur Gravitationsphysik,
Albert-Einstein-Institut, Am M\"uhlenberg 1,
D-14476 Golm, Germany}
\date{Feb 6, 2005}
\email{Reinhard.Prix@aei.mpg.de}

\keywords{multi-fluid hydrodynamics, convective variational
    principle, conducting fluids, plasma hydrodynamics, superconductivity}

\begin{abstract}
In this work we extend our previously developed formalism of
Newtonian multi-fluid hydrodynamics to allow for coupling between the
fluids and the electromagnetic and gravitational field. This is
achieved within the convective variational principle by using a
standard minimal coupling prescription.
In addition to the conservation of total energy and momentum, we
derive the conservation of \emph{canonical} vorticity and
helicity, which generalize the corresponding conserved quantities of
uncharged fluids. 
We discuss the application of this formalism to electrically
conducting systems, magnetohydrodynamics and superconductivity. 
The equations of electric conductors derived here are more general
than those found in the standard description of such systems, in which 
the effect of entrainment is overlooked, despite the fact that
it will generally be present in any conducting multi-constituent
system. 
\end{abstract}

\keywords{multi-fluid hydrodynamics; convective variational principle;
  conducting fluid; MHD; superconductivity}

\pacs{47.10.+g, 47.65.+a, 74.20.-z, 52.30.-q}

\maketitle 


\section{Introduction}

In the previous paper \cite{prix02:_variat_I}, henceforth referred to
as Paper~I, we have shown how the convective variational principle can
be used to derive the general equations of motion for a system of
interacting uncharged fluids. This variational principle is also 
at the heart of a series of papers by Carter and 
Chamel~\cite{carter04:_covar_newtonI,carter03:_covar_newtonII,carter04:_covar_newtonIII} 
based on a fully covariant spacetime formulation of Newtonian
hydrodynamics, which is formally closer to the relativistic formulation.

Here we extend our ``3+1'' framework developed in Paper~I to allow for
charged fluids and their coupling to the electromagnetic and
gravitational field. In Paper~I we already included the gravitational
field via an explicit prescription for the external force, but now
this coupling is derived in a more natural way from the variational
principle using the same minimal coupling procedure as for the
electromagnetic field.

Due to the coupling to the electromagnetic field, the resulting theory
can only be considered an ``approximate'' Newtonian framework, as
strict Galilean invariance will be violated.
While the Newtonian hydrodynamic equations are strictly invariant under Galilean
transformations, the equations of electrodynamics are invariant only
under Lorentz-transformations. This well known discrepancy lead to the
development of relativity, and a coupling between Newtonian physics
and electromagnetism is strictly speaking either inconsistent or
selects a preferred frame (``ether'').  As shown in
\cite{bellac73:_galil_elect}, demanding \emph{strict} Galilean
invariance of electrodynamics without an ether forces one to adopt
either an ``electric'' or ``magnetic'' limit of the theory, in which
certain essential effects would be absent (e.g. the magnetic
force between electric currents or local charge conservation). 
The only fully consistent approach is to work within a (locally)
Lorentz-invariant relativistic framework, as used for example in 
\cite{carter98:_relat_supercond_superfl,carter02:_super_super_neutr_stars}. 

From a practical point of view, however, a ``3+1'' non-relativistic
formalism is often fully sufficient in terms of precision and usually
more easily applicable to many problems. The aim of this paper is
therefore to provide a flexible and general framework for the
description of a wide variety of charged multi-constituent systems in
the ``non-relativistic'' regime of small velocities and low
frequencies $\O\left(c^{-2}\right)$\footnote{to be quantified in the appendix}.  
We emphasize that we do not attempt to construct a strictly
Galilean-invariant theory, rather this should be regarded as 
a truncated theory at order $\O(c^{-2})$ of a Taylor expansion 
of the underlying fully covariant theory.

For simplicity we restrict our analysis to non-magnetic and
non-polarizable fluids, such that the vacuum Maxwell-equations 
retain their form on the macroscopic level of hydrodynamics, and
the interaction between matter and fields is restricted to the minimal
coupling type. The inclusion of electric and magnetic  polarization
is postponed to future work.

Using this variational framework, the fundamental effects
of ``entrainment'' and ``chemical coupling'', inherent to
multi-constituent systems (as discussed in Paper~I), are automatically
included in the formalism. This is a substantial improvement over
the standard ``orthodox'' description of electrical conductivity and
charged fluids in general, in which these effects are usually
completely overlooked. The present framework is also more
general in the sense of being readily applicable to an arbitrary
number of interacting constituents and fluids.

The plan of this paper is as follows:
in Sec.~II we extend the variational principle introduced in Paper~I
to the case of (``separable'') coupling to the electromagnetic and
gravitational fields, and derive the equations of motion for this
coupled system. In Sec.~III we derive the conservation of charge and
mass as well as of energy and momentum from the variational
principle. Sec.~IV is devoted to conserved quantities under transport
of the fluid flow, namely the canonical vorticity and helicity, and
their special relation to superconductors. In Sec.~V we discuss 
applications of this formalism to particular physical systems.
In the appendix we derive the non-relativistic form of electrodynamic
equations and show their ``approximate'' Galilean invariance.

\section{Variational description of multi-constituent systems}
\label{sec:GeneralMultiConst}

\subsection{Kinematics}
\label{sec:Kinematics}

We follow the notation introduced in Paper~I, the various constituents
are indexed by capital letters $\X,\Y,...$, which run over the 
constituents labels. 
The fundamental kinematic variables are the constituent
number-densities $n_\X$ and the associated transport currents
$\vn_\X$, which are related to the respective flow-velocities $\vv_\X$
as $\vn_\X = n_\X \vv_\X$. As in Paper~I, we define the relative
velocities $\vDv_{\X \Y}$ between fluids $\X$ and $\Y$ as
\begin{equation}
  \label{eq:5}
  \vDv_{\X\Y} \equiv \vv_\X - \vv_\Y\,.
\end{equation}
In addition to the masses per particle $m^\X$, we now admit another
transported quantity, namely the charge per particle $q^\X$. 
The corresponding total densities and currents are therefore the 
mass density $\rho$ and mass current $\vec{\rho}$, given by
\begin{equation}
  \label{eq:DefRho}
  \rho = \csum n_\X m^\X\,,\qaq
  \vec{\rho} = \csum m^\X \vn_\X \,,
\end{equation}
and also the charge density $\jo$ and charge current $\vj$, defined as
\begin{equation}
  \label{eq:DefSigJ}
  \jo = \csum n_\X q^\X\,,\qaq
  \vj = \csum \vn_\X q^\X\,,
\end{equation}
where here and in the following $\csum$ is used to indicate the sum
over repeated constituent indices, no automatic summation over
constituent indices is assumed. 

In addition to the dynamics of the matter-system we also want to
include the coupling to the gravitational gauge field $\Phi$ and the
electromagnetic gauge fields $\Ao$ and $\vA$. The corresponding
gauge-invariant field quantities are the gravitational acceleration
$\vec{g}$, which is
\be
\vec{g} = - \vnabla \Phi\,,
\fe
and the electric and magnetic fields $\vE$ and $\vB$, 
which are
\begin{equation}
  \label{eq:DefEMFields}
  \vE \equiv \vnabla \Ao - {1\over c} \partial_t \vA\,,\qaq
  \vB \equiv \vnabla \times \vA\,.
\end{equation}
These fields are invariant under the gauge transformations
\be
\Phi' = \Phi + \mathcal{C}(t)\,,\quad
\Ao' = \Ao + {1\over c}\partial_t \psi\,,\qaq
\vA' = \vA + \vnabla \psi\,.
\fe

\subsection{Dynamics}

The dynamics of the system is described by an action $\Act$ of the form 
\begin{equation}
  \label{eq:DefAct}
  \Act = \int \Lagr \, d V\, d t\,,
\end{equation}
in terms of the (generally gauge-dependent) total Lagrangian $\Lagr$. 
Using the standard minimal coupling prescription, and assuming the
electrodynamic field-contribution to be fully separable from the
hydrodynamic Lagrangian, we can postulate the form of $\Lagr$  to be
\begin{equation}
  \Lagr = \Lagr_\H(n_\X, \vn_\X) +  \Lagr_\EM(\vE, \vB) + \Lagr_\G(\vec{g})
+ (\jo \Ao + {1\over c} \vj\cdot\vA ) - \rho \Phi \,.
\label{eq:TotalLagrangian}
\end{equation}
The gravitational field Lagrangian $\Lagr_\G$ has the well-known expression
\be
\LagrG = -{1\over 8\pi G} (\vnabla \Phi)^2\,,
\fe
where $G$ is Newton's gravitational constant. The electromagnetic
field contribution (cf. appendix) in the non-polarizable case is given
by 
\begin{equation}
  \label{eq:15}
  \Lagr_\EM = {1\over8\pi}\left( \vE^2 - \vB^2 \right)\,.
\end{equation}
As we have seen in the case of uncharged fluids (cf. Paper~I), the
hydrodynamic Lagrangian $\LagrH$ defines the \emph{dynamical} momenta
$p^\X_0$ and $\vp^\X$ per fluid particle as 
\be
d\LagrH = \csum \left(p_0^\X \, d n_\X + \vp^\X\cdot d \vn_\X
\right)\,,
\quad\textrm{so}\quad
p_0^\X = {\partial \LagrH \over \partial n_\X}\,,\quad
\vp^\X = {\partial \LagrH \over \partial \vn_\X}\,.
\label{equVarL}
\fe
In a similar manner the variation of the electrodynamic field
Lagrangian $\Lagr_\EM$ defines the so-called ``electric displacement
field'' $\vD$ and the ``magnetic field strength'' $\vH$ as conjugate
variables to the electromagnetic fields, namely
\be
d \Lagr_\EM = {1\over 4\pi}\vD\cdot d\vE - {1\over 4\pi} \vH\cdot d\vB\,,
\label{eq:dLagrF}
\fe
and using the explicit form (\ref{eq:15}) we find
\be
\vD = 4\pi{\partial \LagrF \over \partial \vE} = \vE\,,\quad 
\vH = - 4\pi {\partial \LagrF \over \partial \vB} = \vB\,.
\label{eq:HD}
\fe
Although we can trivially identify $\vD=\vE$ and $\vH=\vB$ in the
present non-polarizable case, in the following we nevertheless keep
the formal and conceptual separation between the ``kinematical'' $\vE$
and $\vB$ and the  ``dynamical'' (i.e. derived from $\Lagr$) quantities $\vD$ and $\vH$.

Due to the presence of the minimal coupling terms in
(\ref{eq:TotalLagrangian}), the variation of the total Lagrangian
$\Lagr$ generalizes the dynamical momenta $p_0^\X$ and $\vp^X$ to
the (gauge-dependent) \emph{canonical} momenta $\pi^\X_0$ and $\vpi^\X$, namely  
\begin{equation}
  \label{eq:DefCanonicalMomenta}
  d\Lagr = \csum (\pi^\X_0 \, d n_\X + \vpi^\X\cdot d \vn_\X) 
  -\rho \, d\Phi + \jo \,d\Ao + {1\over c}\vj\cdot d \vA + d\LagrF + d\LagrG\,,
\end{equation}
from which we obtain the relations
\bea
  \pi^\X_0 &=&  p^\X_0 + q^\X \Ao - m^\X \Phi \,, \label{eq:RelationDynCanMomenta1}\\
  \vpi^\X &=& \vp^\X + {1\over c} q^\X \vA\,,\label{eq:RelationDynCanMomenta2}
\fea
expressing the canonical momenta in terms of the dynamical momenta and
the gauge fields.

\subsection{The equations of motion}

We see from definition (\ref{eq:DefEMFields}) that the first
two Maxwell equations are satisfied identically, i.e. 
\begin{equation}
  \vnabla \cdot \vB = 0\,,\qaq
  \vnabla\times \vE + {1\over c} \partial_t \vB = 0\,.
  \label{eq:Maxwell1}
\end{equation}
Using (\ref{eq:dLagrF}) and (\ref{eq:DefCanonicalMomenta}) it is not
difficult to show that requiring invariance of the action
(\ref{eq:DefAct}) under free variations $\d\Ao$ and $\d\vA$ of the 
gauge fields results in the two remaining Maxwell equations
\begin{equation}
  \vnabla \cdot \vD = 4\pi \jo\,,\qaq
\vnabla\times \vH - {1\over c} \partial_t \vD = {4\pi \over c} \vj\,.
\label{eq:Maxwell2}
\end{equation}
Similarly, requiring invariance of the action (\ref{eq:DefAct}) with respect to
free infinitesimal variations $\d\Phi$ straightforwardly results in
the Poisson equation for the gravitational field,
\begin{equation}
  \nabla^2 \Phi = 4\pi G \rho\,.
\end{equation}

The derivation of the equations of motion for the fluids
is formally completely equivalent to
the uncoupled case discussed in Paper~I. The only difference is that
the dynamical momenta $p_0^\X$ and $\vp^\X$ are now replaced
by the corresponding canonical momenta $\pi_0^\X$ and $\vpi^\X$. 

One has to consider infinitesimal spatial displacements $\vxi_\X$ and
time-shifts $\dtau_\X$ of the flowlines, which should be considered
the ``true'' fundamental quantities describing hydrodynamics
(corresponding to the Lagrangian framework), while free variations for
densities and velocities (characteristic for the Eulerian framework)
fail to produce the correct equations, except when adding ``ad-hoc''
constraints to the Lagrangian. The Eulerian variables, density $n_\X$
and velocity $\vv_\X$ can be expressed in terms of the underlying
flowlines (and initial conditions) alone, and one should therefore
also consider the hydrodynamic Lagrangian $\LagrH$ in
(\ref{eq:TotalLagrangian}) as a function of these variables. These
subtleties relating to the hydrodynamic variational principle are
discussed in greater detail in Paper~I. 

The resulting induced variations of the densities, $\d n_\X$, and
currents, $\d \vn_\X$ have been derived in Paper~I as 
\bea
\d n_\X &=& - \vnabla\cdot\left[n_\X\vxi_\X \right] 
+\left[ \vn_\X\cdot\vnabla\dtauA - \dtauA\partial_t n_\X\right ]\,,
\label{equdDens}\\
\d\vn_\X &=& n_\X\partial_t \vxi_\X + (\vn_\X\cdot\vnabla)\, \vxi_\X
- (\vxi_\X\cdot\vnabla)\vn_\X - \vn_\X (\vnabla\cdot\vxi_\X) -
\partial_t \left( \vn_\X \dtauA\right)\,.
\label{equdCur}
\fea
Substituting these expressions into the variation of the total Lagrangian
(\ref{eq:DefCanonicalMomenta}) and integrating by parts, we arrive at
the following form of the induced action variation 
\be
\d\Act = \int \csum \left(  g^\X\, \dtau_\X  - \vcf^\X\cdot\vxi_\X
\right)\,d V\,d t\,, \label{equActionVar} 
\fe
where the \emph{canonical} force densities $\vf^\X$ (acting {\em on}
the constituent) and the energy transfer rates $g^\X$ ({\em into} the
constituent) are found explicitly as 
\bea
\vcf^\X &=& n_\X \left(\partial_t \vpi^\X - \vnabla \pi^\X_0 \right) -
\vn_\X \times (\vnabla\times\vpi^\X) + \vpi^\X \,\Rate_\X \,,
\label{equfA}\\
g^\X &=& \vv_\X\cdot\left(\vcf^\X - \vpi^\X \Rate_\X \right) -
\pi^\X_0 \Rate_\X\,,
\label{equgA}
\fea
where $\Rate_\X$ is the particle creation rate for the constituent $\X$, i.e. 
\be
\Rate_\X \equiv  \partial_t n_\X + \vnabla\cdot \vn_\X\,. \label{eq:DefRateX}
\fe
The canonical force density $\vf^\X$ is the total (canonical)
momentum change rate of the constituent $\X$, and the last term in
(\ref{equfA}) represents a contribution that is purely due to the
change of the particle number. 
By inserting the explicit expressions
(\ref{eq:RelationDynCanMomenta1},\ref{eq:RelationDynCanMomenta2}) of
the canonical momenta into (\ref{equfA}) and (\ref{equgA}), we can
separate out the purely hydrodynamic contribution $\vfH^\X$, which
expresses fluid inertia and pressure, and which has the same
form as in the uncharged case (Paper~I), namely
\begin{equation}
  \label{eq:fhydro}
  \vfH^\X \equiv n_\X \left(\partial_t \vp^\X - \vnabla p^\X_0 \right)
  - \vn_\X\times(\vnabla\times \vp^\X) \,.
\end{equation}
Introducing the gravitational and electromagnetic forces $\vfG^\X$ and
$\vfF^\X$, defined by their usual expressions  
\bea
  \vfG^\X &=& - n_\X m^\X \vnabla \Phi \,, \label{eq:Fgrav}\\
  \vfF^\X &=&   n_\X q^\X \left( \vE + {1\over c}\vv_\X \times \vB\right)\,,\label{eq:Fem}
\fea
we can now rewrite the canonical force densities (\ref{equfA}) in the form
\begin{eqnarray}
\vf^\X &=& \vfH^\X - \vfG^\X - \vfF^\X + \Rate_\X \, \vpi^\X \,. \label{eq:cfX}
\end{eqnarray}
The actual physical equations of motion are
obtained by prescribing the canonical force densities $\vf^\X$ acting
on the fluids.
If we only require invariance of the action for a \emph{common}
displacement and time shift $\vxi_\X=\vxi$ and $\dtau_\X=\dtau$, we obtain the minimal equations of motion
for the total system, namely
\be
\csum \vcf^\X = \vcf_\ext\,,\qaq
\csum g^\X = g_\ext\,, \label{eq:GeneralEOM}
\fe
where $\vcf_\ext$ and $g_\ext$ are interpretable as the ``external''
force density and energy transfer rate acting on the system. This
generalizes the more common action principle of \emph{isolated}
systems, in which the external influences $\vcf_\ext$ and $g_\ext$
vanish and therefore the equations of motion would be obtained by
requiring the action to be \emph{invariant} under small variations. 
``External'' here is meant in the sense of not being included in the
total Lagrangian, which could also include, for example, viscous forces. 

\subsection{The hydrodynamic Lagrangian $\LagrH$}

As shown in Paper~I, the hydrodynamic Lagrangian density $\LagrH$ for
the class of ``perfect'' (i.e. with an isotropic energy function $\E$)
multi-fluid systems is given by 
\begin{equation}
  \LagrH(n_\X, \vn_\X) = \csum m^\X {\vn_\X^2 \over 2 n_\X}
  - \E(n_\X, \Dv_{\X\Y}^2)\,,
  \label{equLH}
\end{equation}
where $\vDv_{\X\Y}$ denotes the relative velocity between fluids $\X$
and $\Y$ as defined in (\ref{eq:5}). The total differential of the
thermodynamic potential $\E(n_\X, \Dv_{\X\Y}^2)$ determines the first
law of thermodynamics for the matter-system, namely 
\be
d\E =  \csum \mu^\X \,d n_\X + 
{1\over2} \csum_{\X,\Y}\entr^{\X\Y} \, d\Dv^2_{\X\Y}\,, 
\label{equdE}
\fe
which defines the chemical potentials $\mu^\X$ and the symmetric
entrainment matrix $\entr^{\X\Y}$.
The dynamical momenta $p^\X_0$ and $\vp\,^\X$ defined in
(\ref{equVarL}) are therefore found as    
\bea
\vp^\X &=& m^\X \vv_\X - \csum_\Y {2\entr^{\X\Y}\over n_\X}
\vDv_{\X\Y}\,, \label{equpA}\\ 
-p^\X_0 &=& \mu^\X -  m^\X {v_\X^2 \over 2}  +
\vv_\X\cdot\vp^\X\,. \label{equp0A} 
\fea

\section{Gauss-type conservation laws}

\subsection{Conservation of charge and mass}

We see from (\ref{eq:fhydro})--(\ref{eq:Fem}) that the force
contributions  $\vfH^\X$, $\vfG^\X$ and $\vfF^\X$ are invariant  
with respect to gauge transformations of the gravitational and
electromagnetic fields. However, the last term in the expressions
(\ref{equfA}) and (\ref{equgA}) (accounting for momentum and energy
change due to particle number changes) is generally gauge dependent. 
While this is not prohibited for individual constituent forces, the 
total equations of motion (\ref{eq:GeneralEOM}) have to be gauge
invariant, and by using
(\ref{eq:RelationDynCanMomenta1},\ref{eq:RelationDynCanMomenta2}) we
can therefore deduce the constraints   
\begin{eqnarray}
  \csum q^\X \Rate_\X &=& \partial_t \jo + \vnabla\cdot \vj = 0\,,\label{eq:ChargeCons}\\
  \csum m^\X \Rate_\X &=& \partial_t \rho + \vnabla\cdot \vec{\rho} =
  0\,,\label{eq:MassCons}
\end{eqnarray}
where the total densities and currents have been defined in
(\ref{eq:DefRho}) and (\ref{eq:DefSigJ}). Gauge invariance therefore
implies conservation of the ``charge'' associated with the gauge field.
As a further consequence we find the following useful relations,
\begin{equation}
  \csum \Rate_\X \vpi^\X = \csum \Rate_\X \vp^\X\,,\qaq
  \csum \Rate_\X \pi^\X_0 = \csum \Rate_\X p^\X_0\,.
  \label{eq:CsumPi}
\end{equation}
We note that the conservation of mass was already derived in Paper~I
as a consequence of Galilean invariance, which can also be considered
a gauge freedom. 

\subsection{Momentum conservation}
\label{secMomCons}

As shown in Paper~I, the purely hydrodynamic force densities $\vfH^\X$
satisfy the relation
\be
\csum (f_\H^{\X\,i} + \Rate_X \,p^{\X\,i}) = \partial_t J_\H^i +
\nabla_j T_\H^{i j}\,,
\fe
where the hydrodynamic momentum density $\vJ_\H$ and stress tensor
$T_\H^{i j}$ are given by 
\begin{equation}
  \label{eq:Tmunu}
  \vJ_\H \equiv \csum n_\X \vp^\X\,,\qaq 
T_\H^{i j} \equiv \csum n_\X^i p^{\X\,j} + \Press\, g^{i j}\,.
\end{equation}
The ``generalized pressure'' $\Press$ is defined via the Legendre
transformation of $\LagrH$, namely 
\be
\Press \equiv \LagrH - \csum \left( n_\X p_0^\X + \vn_\X\cdot \vp^\X\right)\,,
\label{eq:DefPress}
\fe
and $g_{i j}$ are the components of the metric tensor, which in
Cartesian coordinates is simply $g_{i j}=\delta_{i j}$. 
Using these relations together with (\ref{eq:cfX}) and
(\ref{eq:CsumPi}), we can write the total force balance equation  
(\ref{eq:GeneralEOM}) in the form 
\begin{equation}
  f_\ext^i = \csum f^{\X\,i} = \partial_t J_\H^i + \nabla_j T_\H^{i j}\
  - f_\G^i - f_\EM^i\,,
\label{eq:ForceBalance}
\end{equation}
where \mb{$\vfG\equiv \csum \vfG^\X$} and \mb{$\vfF\equiv \csum \vfF^\X$}
are the total gravitational and electromagnetic force densities. 
Using (\ref{eq:Fgrav}) and (\ref{eq:Fem}) we find explicitly
\bea
  \vfG &=& -\rho \vnabla \Phi\,,\\
  \vfF &=& \jo \vE + {1\over c} \vj \times \vB\,. \label{eq:EMForce}
\fea
One can easily verify that the gravitational force term can be written
as the divergence of a tensor, namely 
\begin{equation}
-f_\G^i = \nabla_j T_\G^{i j}\,,\quad\textrm{with}\quad
  T_\G^{i j} = {1\over 4\pi G}\left( \nabla^i \Phi \nabla^j \Phi -
   {1\over 2}(\vnabla \Phi)^2 \,g^{i j} \right)\,.
 \label{eq:Tgrav}
\end{equation}
Using the Maxwell equations (\ref{eq:Maxwell1}), (\ref{eq:Maxwell2})
and the total differential (\ref{eq:dLagrF}) of $\Lagr_\EM$, we can
show that the total electromagnetic force density (\ref{eq:EMForce})
can be similarly rewritten as   
\begin{equation}
-f_\EM^i = \partial_t \Jem^i + \nabla_j T_\EM^{i j}\,, 
  \label{eq:EMForceCons}
\end{equation}
in terms of the momentum density $\vec{J}_\EM$ of the electromagnetic
field, 
\begin{equation}
  \vec{J}_\EM \equiv {1\over 4\pi c} \vD\times \vB\,,
  \label{eq:Poynting}
\end{equation}
and the Maxwell stress tensor $T_{\EM}^{i j}$, which is found as
\begin{equation}
  T_\EM^{i j} = -{1\over 4\pi}\left( E^i D^j + H^i B^j\right)
+ \left( \LagrF + {1\over 4\pi}\vH\cdot\vB \right)\, g^{i j} \,.
\label{eq:Tem}
\end{equation}
Putting all the pieces together, we obtain the following form for the
total momentum conservation (\ref{eq:ForceBalance}):
\begin{equation}
  \label{eq:TotalMomCons}
  \partial_t (J_\H^i + \Jem^i) + \nabla_j T^{i j} = f_\ext^i\,,
\end{equation}
where the total stress tensor is given by
\begin{equation}
  \label{eq:TotalStress}
  T^{i j} \equiv T_\H^{i j} + T_\G^{i j} + T_\EM^{i j}\,.
\end{equation}

An important property of the \emph{total} stress tensor $T^{i j}$  is
that it is symmetric. The symmetry of the gravitational part
(\ref{eq:Tgrav}) is obvious, while the symmetry of the hydrodynamic
stress tensor (\ref{eq:Tmunu}) has been shown in Paper~I. 
It remains to prove the symmetry of the electromagnetic stress tensor
(\ref{eq:Tem}). In the present non-polarizable case, this follows
trivially from the identities (\ref{eq:HD}). It is interesting to
note, however, that this can also be derived more generally as a
Noether identity of the variational principle, assuming only the
separable form (\ref{eq:TotalLagrangian}) of the Lagrangian.  

In order to show this, we extend the variational principle slightly by
admitting also metric variations $\d g_{i j}$, so that (\ref{eq:dLagrF}) now reads as
\be
\d \Lagr_\EM = {1\over 4\pi} \vD\cdot\d\vE - {1\over 4\pi}\vH\cdot\d\vB
+ {\partial \Lagr_\EM \over \partial g_{i j}}\, \d g_{i j}\,.
\label{eq:dMetricLagrF}
\fe
Consider an \emph{active} time-independent infinitesimal displacement $\vxi$
of the whole system including the metric, which induces the following
Lagrangian changes:
\be
\D E_i = E^l\nabla_i \xi_l \,,\quad
\D B_i = B^l\nabla_i \xi_l \,,\quad
\D g_{i j} = - 2 \nabla_{(i} \xi_{j)}\,.
\fe
Using these transformations together with (\ref{eq:dMetricLagrF}), we
obtain the induced Lagrangian change of $\Lagr_\EM$ as
\be
\D \Lagr_\EM = {1\over 4 \pi}\left[ D^i \,E^j - H^i \,B^j 
- 8\pi {\partial \Lagr_\EM \over \partial g_{i j} } \right]\, \nabla_i \xi_j\,.
\fe
This active transformation is equivalent to a coordinate
transformation $-\vxi$, and therefore the requirement of $\Lagr_\EM$
being a scalar is \mb{$\D \Lagr_\EM = 0$}, which leads to the
associated Noether identity  
\be
2 {\partial \Lagr_\EM \over \partial g_{i j}} = 
 {1\over 4 \pi}\left[ D^i \,E^j - H^i \,B^j \right]\,.
\fe
Using the manifest symmetry of the left-hand side, we obtain
\be
E^i \, D^j + H^i \, B^j = E^j \, D^i + H^j \, B^i\,,
\fe
which concludes the proof of the symmetry of $T_\EM^{i j}$.
Note that in general $T_\EM^{i j}$ need not be symmetric, only the sum of
all contributions to $T^{i j}$ is subject to this constraint. The
symmetry of $T_\EM^{i j}$ is a special consequence of the assumption
of a ``separable'' interaction of the form (\ref{eq:TotalLagrangian}).

\subsection{Energy conservation}

We have seen in Paper~I that we can write
\begin{equation}
  \csum \left(\vv_\X\cdot\vfH^\X - \Rate_\X p^\X_0 \right) =
  \partial_t E_\H + \vnabla\cdot\vQ_\H\,, 
\end{equation}
in terms of the hydrodynamic energy density $E_\H$ and energy flux
$\vQ_\H$, which are given by 
\be
  \label{eq:T0mu}
  E_\H = \csum \vn_\X\cdot\vp^\X - \LagrH\,,\qaq
  \vQ_\H = \csum (-p^\X_0)\, \vn_\X\,,
\fe
while for the gravitational and electromagnetic work contributions
\mbox{$g_\G\equiv\csum \vv_\X \cdot \vfG^\X$} and 
\mbox{$g_\EM \equiv \csum \vv_\X \cdot \vfF^\X$}, we find using
(\ref{eq:Fgrav}) and (\ref{eq:Fem}):
\bea
g_\G &=& - \vec{\rho}\cdot\vnabla\Phi \,,\\
g_\EM &=&  \vj\cdot\vE\,.
\fea
Using these expressions, the energy equation (\ref{eq:GeneralEOM})
can be written as
\begin{equation}
  \label{eq:EnergyCons2}
  g_\ext =  \partial_t E_\H + \vnabla\cdot\vQ_\H - g_\G - g_\EM\,.
\end{equation}
Using Maxwell's equations (\ref{eq:Maxwell1}) and (\ref{eq:Maxwell2}),
one can write the electric work $g_\EM$ in the form of a conservation
law, namely 
\begin{equation}
  -g_\EM = \partial_t E_\EM + \vnabla\cdot \vS\,,
\end{equation}
where the electromagnetic field energy density $E_\EM$ is given by
\begin{equation}
  E_\EM = {1\over 4\pi} \vE\cdot\vD - \LagrF = {1\over 8\pi}\left(\vE^2+\vB^2 \right)\,,
\end{equation}
and the second equality was obtained using the explicit Lagrangian (\ref{eq:15}).
The energy flux $\vS$ is given by the Poynting vector 
\be
\vS = {c\over 4\pi}\, \vE \times \vH\,.
\fe
In the present non-polarizable case, i.e. \mb{$\vD=\vE$} and
{$\vB=\vH$}, we recover the well-known relation between the
energy flux and momentum of the electromagnetic field, namely
\mb{$\vec{J}_\EM = \vS/c^2$}.
Summarizing, we can cast (\ref{eq:EnergyCons2}) in the form of a
conservation of total energy, namely
\begin{equation}
  \label{eq:EnergyCons}
  \partial_t (E_\H + E_\EM) + \vnabla\cdot(\vQ_\H +
  \vS) = g_\ext - \vec{\rho}\cdot\vnabla\Phi\,.
\end{equation}
We note that formally one can also write the gravitational work in the
form of a conservation law, but the expression for energy density and
flux are neither unique nor gauge invariant, and one can also not
eliminate the mass current $\vec{\rho}$ from these expressions due to
the lack of a dynamic law for the gravitational field in the Newtonian
framework. 

\section{Conservation along flowlines}
\label{sec:FlowlineConservations}

In this section we show how the conservation of vorticity and helicity,
derived for uncharged fluids in Paper~I, can be generalized quite
naturally to the case of fluids coupled to the electromagnetic and
gravitational field. We note that the technical steps involved in this
discussion are largely analogous to the treatment in Paper~I,
and we therefore skip most intermediate steps.

\subsection{Generalized Kelvin-Helmholtz vorticity conservation}
\label{sec:Vorticity}

We define the hydrodynamic vorticity 2-form $\form{\vort}$
as the exterior derivative of the dynamical momentum 1-form $\form{p}$, namely 
$\form{\vort}\equiv d\form{p}$, and the more common dual vorticity
\emph{vector} $\vec{\Vort}$, which is $\vec{\Vort} = \vnabla \times \vp$.

In the presence of electromagnetic fields, the more fundamental quantity
is the \emph{canonical} vorticity 2-form $\form{\cvort}$, which is
defined in the same way but with respect to the canonical momentum $\form{\pi}$, namely  
\begin{equation}
  \label{eq:Defcvort}
  \form{\cvort} \equiv d \form{\pi}\,,
\end{equation}
and the dual canonical vorticity vector $\vec{\cVort}$ is therefore
given by
\begin{equation}
  \vec{\cVort} = \vnabla \times \vpi\,.
\end{equation}
With (\ref{eq:RelationDynCanMomenta2}) we see that the relation
between canonical and hydrodynamic vorticity is simply 
\be
  \label{eq:RelVortCVort}
  \form{\cvort} = \form{\vort} + {q\over c} \, d\form{A}\,,\qaq
  \vec{\cVort} = \vec{\Vort} + {q\over c} \,\vB\,.
\fe
We note that by the Poincar\'e property (namely \mb{$d d = 0$}), the
exterior derivatives of the vorticity 2-forms vanish identically, i.e.
$d \form{\cvort} = 0$, which equivalently expresses the fact that the
vorticity vectors are divergence-free,
i.e. \mb{$\vnabla\cdot\vec{\cVort} = 0$}. 

We can write the expression (\ref{equfA}) for the canonical force $\vf$
acting on one constituent in the language of forms as
\begin{equation}
  \partial_t \form{\pi} + \vv\rfloor d\form{\pi} - d \pi_0 = {1\over
  n}( \form{\cf} - \Rate \form{\pi})\,, 
  \label{eq:forceq1}
\end{equation}
where $\rfloor$ indicates summation over adjacent vector- and form-indices.

In the following it will be convenient to separate the ``proper
force'' per particle acting on the right-hand side of
(\ref{eq:forceq1}) into its non-conservative part $\form{\FF}$ and a
conservative contribution $d\phi$, namely 
\begin{equation}
  {1\over n}( \form{\cf} - \Rate \form{\pi}) = d\phi + \form{\FF}\,.
  \label{eq:ForceSeparation}
\end{equation}
Applying the Cartan formula for the Lie derivative of a $p$-form to
the  1-form $\form{\pi}$, namely
\mbox{$\Lie_\vv\, \form{\pi} = \vv \rfloor d\form{\pi} + d(\vv \rfloor \form{\pi})$},
allows us now to rewrite the force equation (\ref{eq:forceq1}) more
conveniently as   
\begin{equation}
  (\partial_t + \Lie_\vv)\, \form{\pi} = d Q + \form{\FF}\,,
  \label{eq:Helmholtz1}
\end{equation}
where the scalar $Q$ is given by $Q = \pi_0 +   \vv\rfloor\form{\pi} + \phi$.
Lie derivatives and partial time derivatives commute with exterior
derivatives, so we can apply an exterior derivative to
(\ref{eq:Helmholtz1}) and with (\ref{eq:Defcvort}) we obtain  the
Helmholtz equation of transport of canonical vorticity, namely  
\begin{equation}
  \label{eq:HelmholtzIa}
  (\partial_t + \Lie_\vv) \,\form{\cvort} = d \form{\FF}\,,
\end{equation}
which shows that the canonical vorticity is conserved under transport
by the fluid, if the proper force per particle acting on the fluid is 
purely conservative, i.e. if \mb{$\form{\FF} = 0$}.
In its more common dual form, this equation can be written as
\begin{equation}
  \label{eq:HelmholtzIb}
  \partial_t \vec{\cVort} - \vnabla\times\left(\vv \times
  \vec{\cVort}\right) = \vnabla\times \vec{\FF}\,,
\end{equation}
Substituting the explicit relation (\ref{eq:RelVortCVort}) between the
 dynamical and the canonical vorticity, and using Maxwell's equation,
 this can be re-expressed as 
\be
\partial_t \vec{\Vort} - \vnabla \times (\vv \times \vec{\Vort}) = 
\vnabla \times \left[\vec{\FF} + q \left(\vE +{1\over c}\vv\times \vB\right) \right]\,,
\fe
which shows that in the case of a charged constituent, the dynamical
vorticity $\vec{\Vort}$ is generally not conserved even in the
absence of a non-conservative external force $\vec{\FF}$, due to the
presence of the Lorentz force acting on the fluid. However, the
 canonical vorticity \emph{is} conserved in this case and therefore
 generalizes the vorticity conservation of uncharged fluids.

For the canonical circulation $\Circ$ of a closed circuit
$\partial\Sigma$, which is the boundary of a 2-surface $\Sigma$, we
have  
\begin{equation}
  \label{eq:DefCirculation}
  \Circ \equiv \oint_{\partial\Sigma}\form{\pi} = \int_{\Sigma}
  \form{\cvort} = \int_{\Sigma} \vec{\cVort}\cdot d\vec{S}\,,  
\end{equation}
where $d\vec{S}$ is the surface normal element.
We see from (\ref{eq:RelVortCVort}) that the canonical circulation
$\Circ$ can also be expressed as the sum of the hydrodynamic vorticity
flux (i.e. dynamical circulation) and the magnetic flux through the
surface $\Sigma$, namely  
\begin{equation}
  \Circ = \int_{\Sigma} \vec{\Vort}\cdot d\vec{S} + {q\over c}
  \int_\Sigma \vB\cdot d\vec{S}\,.
\end{equation}
For the comoving time derivative of the circulation $\Circ$ we find
using (\ref{eq:Helmholtz1}) 
\begin{eqnarray}
  {d \Circ \over d t}&=& \oint (\partial_t + \Lie_\vv) \, \form{\pi} =
  \oint_{\partial\Sigma} \form{\FF}
  \label{eq:Kelvin}
\end{eqnarray}
which is Kelvin's theorem for the conservation of canonical circulation.
We note that strict conservation only applies if the non-conservative
force per particle $\form{\FF}$ vanishes, as we have already seen
earlier.

\subsection{Vorticity and superconductors}
\label{sec:Superfluidity}

As discussed in more detail in Paper~I, the hydrodynamics of
superfluids is generally characterized by two fundamental properties:
the absence of dissipative mechanisms like friction or viscosity, and
the constraint of irrotational flow. While in the case of uncharged
superfluids this simply meant the vanishing of the dynamical vorticity
$\vec{\Vort}$, it is now the \emph{canonical} vorticity $\vec{\cVort}$
that is constrained to vanish identically in the case of charged
superfluids, more commonly referred to as ``superconductors''.
The absence of microscopic dissipative mechanisms implies that
there is no non-conservative force acting on the bulk\footnote{As
  mentioned in Paper~I, this condition can be violated in the core of
  vortices, leading to ``mutual friction''.} of the superfluid, i.e.
\begin{equation}
\vec{\FF}^\S =0\,,
\label{eq:SfForceFree}
\end{equation}
which quite generally characterizes perfect conductors of any sort.
As a consequence we see that the canonical vorticity (and
equivalently circulation) of a perfect conductor is strictly
conserved, as seen in the previous section. The further constraint of
irrotational flow, which distinguishes a superconductor from a mere
perfect conductor, reads as 
\begin{equation}
  \label{eq:IrrotFlow}
  \form{\cvort}^\S = \form{\vort}^\S + {1 \over c} q^\S\, d\form{A} = 0\,,\qaq
  \vec{\cVort}^\S = \vec{\Vort}^\S + {1\over c}q^\S \vB = 0\,.
\end{equation}
We see from (\ref{eq:HelmholtzIa}) or (\ref{eq:HelmholtzIb}) that if
this irrotationality constraint is satisfied at some instant $t$, then it
will automatically remain true for all subsequent times due to the
absence of dissipation (\ref{eq:SfForceFree}).
We can therefore write the superfluid momentum $\form{\pi}^\S$
(locally) as the gradient of a phase $\varphi$, i.e.
\begin{equation}
  \vpi^\S = \vp^\S + {q^\S\over c}\vA = \hbar\, \vnabla \varphi\,,
\end{equation}
which leads to the well-known London equation for superconductors, as
further discussed in Sect.~\ref{sec:sc}. 
The canonical circulation (\ref{eq:DefCirculation}) can therefore be
non-zero if $\partial\Sigma$ encloses a topological defect in the
phase $\varphi$, i.e. a region where $\varphi$ (and therefore
$\vpi^\S$) is not defined, as for example in the case of flow inside a
torus, or around a vortex. While in the case of a perfect irrotational
fluid the resulting circulation could have any value, the superfluid
phase $\varphi$ is restricted to change by a multiple of $2\pi$ when
following a closed loop inside the superfluid around the defect. The
resulting canonical circulation is therefore quantized as 
\begin{equation}
  \Circ = 2 N \pi \hbar \,,\quad \textrm{with}\quad N \in \mathbb{Z}\,,
\end{equation}
which gives rise to the well-known quantized vortex structure and flux
quantization of superconductors.

\subsection{Generalized helicity conservation}

We now turn to the generalization of the dynamical helicity
conservation derived in Paper~I. We define the canonical helicity
3-form $\form{H}$ as the exterior product of the momentum 1-form
$\form{\pi}$ with the vorticity 2-form $\form{\cvort}$, i.e.  
\begin{equation}
  \label{eq:DefHelicity3Form}
  \form{H} \equiv \form{\pi}\wedge\form{\cvort}\,,
\end{equation}
and we define the dual canonical helicity density $h$ as
\begin{equation}
  \label{eq:HelicityScalar}
  \form{H} = h \, \form{\eps},
\end{equation}
where $\form{\eps}$ is the volume form with components $\eps_{ijk}$.
The helicity scalar can be seen to have the following explicit expressions
\begin{equation}
  h = \form{\pi}\,\rfloor\vec{\cVort} = \vpi\cdot(\vnabla\times\vpi)\,. 
  \label{eq:ExplHel}
\end{equation}
Using (\ref{eq:Helmholtz1}) and  (\ref{eq:HelmholtzIa}), the comoving
time-derivative of $\form{H}$ can be found as 
\begin{eqnarray}
  (\partial_t + \Lie_\vv)\, \form{H} &=& d (Q \form{\cvort}) 
  + \left[  d(\form{\pi} \wedge \form{\FF}) + 2d\form{\FF}\wedge \form{\pi}\right] \,.
\end{eqnarray}
If we further introduce the total canonical helicity $\Hel$ of a volume $V$ as
\begin{equation}
  \label{eq:TotalHelicity}
  \Hel \equiv \int_V \form{H} = \int_V h \, d V\,,
\end{equation}
then we find in the absence of non-conservative forces,
i.e. \mb{$\form{\FF}=0$}, that the comoving time derivative of $\Hel$
satisfies  
\begin{equation}
  \label{eq:HelicityConservation}
  {d \Hel \over d t} = \oint_{\partial V} Q \vec{\cVort} \cdot d\vec{S}\,. 
\end{equation}
Therefore the canonical helicity $\Hel$ of a volume $V$ is conserved
under transport by the fluid only if, in addition to $\form{\FF}=0$,
the canonical vorticity $\vec{\cVort}$ vanishes on the surface
$\partial V$ surrounding this volume.  
We note that in general the conserved helicity $\Hel$ contains
contributions from the purely hydrodynamic ``Moffat'' helicity
$\vp\cdot\vec{\Vort}$ and the magnetic helicity $\vA\cdot\vB$ together
with ``mixed'' terms, namely using (\ref{eq:ExplHel}) we can express
\begin{equation}
  h = \vp\cdot\vec{\Vort} + {q^2 \over c^2} \vA\cdot\vB + {q\over
  c}\left[ \vp\cdot \vB + \vA \cdot \vec{\Vort}\right]\,. 
\end{equation}

\section{Applications}
\label{sec:Applications}

\subsection{General description of electric conductors}

As a simple application of the foregoing formalism, we consider an
electric conductor describable as a two-constituent system. One
constituent consists of the positively charged ions, described by
their number density $n$, velocity $\vv$, mass per ion $m$ and charge
per ion $q=Z e$. The second constituent is a gas of electrons of
density $n_\e$, velocity $\vv_\e$, mass $m_\e$ and charge
$q^\e=-e$. The total charge density and current (\ref{eq:DefSigJ}) are
therefore expressible as   
\begin{equation}
  \jo = e ( Z n - n_\e )\,,\qaq
  \vj = e ( Z n \vv - n_\e \vv_\e)\,,
  \label{eq:ElChargeCurrent}
\end{equation}
and the relative velocity between the two fluids is
\begin{equation}
  \vDv \equiv \vv - \vv_\e\,.
\end{equation}
Charge transfer between the two fluids is possible in principle,
e.g. we could allow for  processes of ionization and recombination,
where electrons are transferred from the ion-fluid to the fluid of
free electrons. But for simplicity we will assume the number of free
electrons to be conserved, so we have 
\begin{equation}
  \Rate = \partial_t n + \vnabla\cdot (n \vv) =0\,,\qaq
  \Rate_e = \partial_t n_\e + \vnabla \cdot(n_\e \vv_\e) = 0\,.
\end{equation}
The total differential of the energy function $\E(n, n_\e, \vDv^2)$ is
\begin{equation}
  d \E = \mu \, d n + \mu^\e \, d n_\e + \entr\, d \Dv^2\,.
  \label{eq:ElConddE}
\end{equation}
Using  (\ref{equpA}) and (\ref{equp0A}), the conjugate momenta of
electrons and ions are therefore found as
\begin{equation}
  \begin{array}{l l}
    \vp = m \vv - {2 \entr \over n} \vDv\,, & -p_0 = \mu - {1\over2} m
    v^2 + \vv\cdot\vp\,,\\
\\
    \vp^\e = m_\e \vv_\e + {2\entr \over n_\e} \vDv \,, & -p^\e_0
    = \mu^\e - {1\over 2} m_\e v_\e^2 + \vv_\e\cdot \vp^\e\,.\\
\label{eq:ElCondEntr}
  \end{array}
\end{equation}
We neglect the gravitational field, so $\vfG=0$, and the canonical
force densities acting on the electron- and ion-fluid are obtained
from (\ref{eq:Fem}) and  (\ref{eq:cfX}) as
\be
  \vcf = \vfH - n Z e ( \vE + {\vv\over c}\times \vB )\,,\qaq
  \vcf^\e = \vfH^\e  + e n_\e (\vE + {\vv_\e\over c}\times \vB ) \,,
\label{eq:210}
\fe
where the hydrodynamic force densities are obtained from
(\ref{eq:fhydro}), by substituting the dynamical momenta
(\ref{eq:ElCondEntr}), which yields
\begin{eqnarray}
 \vfH &=& n m (\partial_t + \vv\cdot\vnabla) \left[
  \vv- {2\entr\over n m }\vDv\right] + n \vnabla \mu - 2\entr \Dv_j
  \vnabla v^j\,, \label{eq:211}\\ 
 \vfH^\e &=& n_\e m_\e (\partial_t + \vv_\e\cdot\vnabla) \left[
  \vv_\e + {2\entr \over n_\e m_\e} \vDv \right] + n_\e \vnabla \mu^\e +
  2\entr \Dv_j \vnabla v_\e^j\, \label{eq:212}.
\end{eqnarray}
These equations contain the description of superconductors,
magneto-hydrodynamic and the fluid-description of plasmas (e.g. see 
\cite{chandrasekhar60:_plasms_physics,freidberg87:_ideal_MHD}) as
special cases. However, they are substantially more general due to the
inclusion of the effect of entrainment, which is usually overlooked in
these contexts.  

Using the momenta (\ref{eq:ElCondEntr}) and the energy differential
\ref{eq:ElConddE}, the generalized pressure  differential
(\ref{eq:DefPress}) is found as
\begin{equation}
  d\Press = n \,d\mu + n_\e \, d \mu^\e - \entr \,d \Dv^2\,.
  \label{eq:213}
\end{equation}
We note that in general we cannot introduce ``partial pressures'',
say, by defining $d P_e$ to be equal $n_\e d\mu^\e$, as this is
generally not a total differential due to interaction energies between
the constituents (i.e. the fact that \mb{$\mu^\e =\mu^\e(n, n_\e,\vDv^2)$}).
However, the chemical potentials are always well-defined and are
therefore much more natural quantities in general multi-fluid contexts.
In the absence of external forces, i.e. \mb{$\vcf_\ext=0$}, the force
balance equation (\ref{eq:ForceBalance}) now reads as 
\begin{equation}
0 = \vcf + \vcf^\e = \vfH + \vfH^\e - \jo\vE - {\vj \over c} \times \vB\,. 
\label{eq:117}
\end{equation}
We can further prescribe a mutual force between the two fluids, so
we introduce a resistivity force of the form \mb{$\vcf^\e = \vec{f}_\R$},
and therefore \mb{$\vcf = - \vec{f}_\R$}.
The energy equation (\ref{eq:GeneralEOM}) with (\ref{equgA}) now takes
the form   
\begin{equation}
  g + g^\e = -\vDv \cdot \vec{f}_\R = g_\ext\,.
\end{equation}
Such a resistive force will lead to creation of heat (entropy), which
in this model has to be extracted to an ``external'' system via
$g_\ext$, as for simplicity we have not explicitly included an entropy
constituent in this example. 
By the second law of thermodynamics, the friction should produce heat
and not absorb it, so we have to extract heat-energy from the system,
i.e. \mb{$g_\ext<0$}, and therefore we can constrain the resistivity
force to be of the form 
\begin{equation}
  \vec{f}_\R = \eta \,\vDv\,,\quad \textrm{with}\quad
  \eta > 0 \,,
  \label{eq:ElFriction}
\end{equation}
where $\eta$ is generally a function of the state-variables describing
the system. 

\subsection{The MHD limit}

In the low-frequency, long-wavelength limit we can assume any
net charge densities to be compensated very quickly by the motion of
electrons, so we make the ``quasi-neutral'' approximation and set:
\begin{equation}
  \jo = 0\,,
\end{equation}
which by (\ref{eq:ElChargeCurrent}) implies \mb{$Z n = n_\e$}, and the
current density therefore reads as
\begin{equation}
  \vj = e n_\e \vDv\,.
  \label{eq:ElCurrent}
\end{equation}
In this low frequency limit we can equally neglect the displacement
current $\partial_t \vD$ in Maxwell's equations (\ref{eq:Maxwell2}).
Because the electrons are very light, i.e. $m_\e \ll m$, the inertial
forces of the electron fluid can usually be neglected as well, and
so the equation of motion for the electrons, \mb{$\vcf^\e = \vcf_\R$},
can be written with (\ref{eq:210}), (\ref{eq:212}) and
(\ref{eq:ElFriction}) as
\begin{equation}
  n_\e \vnabla \mu^\e + 2\entr \Dv_j \vnabla v_\e^j + e n_\e ( \vE +
  {\vv_\e\over c}\times \vB ) = \eta \vDv\,.
  \label{eq:ElTransp}
\end{equation}
In order to recover the ``standard'' MHD framework, we further neglect
entrainment, i.e. if we set \emph{ad hoc} $\entr=0$, so the pressure differential
(\ref{eq:213}) now reduces to 
\begin{equation}
  d P= n_\e \, d \mu^\e + n \, d \mu\,,
\end{equation}
where the generalized pressure $\Press$ can be identified with the
usual pressure $P$ in the absence of entrainment.
This allows us to write the force balance equation (\ref{eq:117}) in
the form
\begin{equation}
  \rho \,(\partial_t + \vv\cdot\vnabla) \vv + \nabla P - {\vj\over
  c}\times \vB = 0\,.
\end{equation}
Using (\ref{eq:ElCurrent}) we can express the electron velocity as
\begin{equation}
  \vv_\e = \vv - {1\over e n_\e} \vj\,,
\end{equation}
and so we can write the equation of electron transport
(\ref{eq:ElTransp}) further as
\begin{equation}
  \vj = \conduct ( \vE + {\vv \over c}\times \vB) + {\conduct\over
  e}\vnabla \mu^\e - {\conduct\over e n_\e c}\vj \times \vB\,,
\end{equation}
where the scalar conductivity $\conduct$ is related to the resistivity
coefficient $\eta$ as
\begin{equation}
  \conduct = {e^2 n_\e^2 \over \eta} > 0\,.
\end{equation}
If we further neglect the ``partial pressure'' $\vnabla \mu^\e$, we can
write the relation between current $\vj$ and electric field $\vE'$ in
the frame of the ion-background, i.e. $\vE'\equiv \vE+(\vv/c)\times\vB$ as a
generalized Ohm's law, namely
\begin{equation}
  j_i = \conduct_{i k} {E^k}' \,,
\end{equation}
where the anisotropic conductivity tensor $\conduct_{i k}$ is 
\begin{equation}
  \conduct_{i k} = \left[ {1\over \conduct}\delta_{i k} +  {1\over e n_\e
  c} \eps_{i k l} B^l\right]^{-1}\,, 
\end{equation}
which is not symmetric but satisfies the relation
\begin{equation}
  \conduct_{i k}(\vB) = \conduct_{k i}(-\vB)\,.
\end{equation}
In this form the generalized Ohm's law can account for the well-known
(classical) Hall effect, while the standard MHD approach (e.g. see
\cite{cowling76:_MHD,jackson75:_class_elect}) commonly also neglects
the ``Hall term'' $\vj\times\vB$, so that this equation finally
reduces to the standard Ohm's law: 
\begin{equation}
  \vj = \conduct \vE' = \conduct ( \vE + {1\over c}\vv\times\vB )\,.
\end{equation}
We note that the ``orthodox'' equations are contained in this
framework as special cases, but the description 
(\ref{eq:ElTransp}) is substantially more general due to the inclusion
of the entrainment effect, which will generally be present in any
(interacting) multi-fluid system.

\subsection{Superconductors}
\label{sec:sc}
In contrast to the previous application, superconductors are perfect
conductors, so the electrons can flow past the ions without friction,
i.e. $\eta=0$ in (\ref{eq:ElFriction}). Therefore we cannot neglect
the inertial and pressure forces of the electrons a-priori.
As mentioned previously (cf. Sect.~\ref{sec:Superfluidity}), 
in addition to the absence of friction, superfluids are also constrained to
be irrotational, so
\begin{equation}
\cvort_{i j} = \nabla^\nix_{[i}\pi^\e_{j]}=0\,,
\label{eq:ScIrrotFlow}
\end{equation}
and in its dual formulation this explicitly reads as
\begin{equation}
\vec{\cVort}^\e =  \vnabla \times \vec{\pi}^\e = \vnabla \times \vp^\e
- {e\over c} \vB = 0\,,  \label{eq:London0a} 
\end{equation}
which we will see after translation to the ``orthodox'' language
represents the (second) London equation. 
In the absence of ``external'' forces acting on the electron fluid,
using (\ref{equfA}) we can reduce the equation of motion for the
electrons, $\vcf^\e=0$, to the form
\begin{eqnarray}
0 &=&  \partial_t \vpi^e - \vnabla \pi^\e_0 = \partial_t \vp^\e -
\vnabla p^\e_0 + e(\vnabla A_0 - {1\over c}\partial_t \vA )\,, \nonumber \\
&=& \partial_t \vp^\e - \vnabla p^\e_0 + e \vE\,, \label{eq:London0b}
\end{eqnarray}
where we have used
(\ref{eq:RelationDynCanMomenta1},\ref{eq:RelationDynCanMomenta2}) and 
the definition (\ref{eq:DefEMFields}) of the electric field $\vE$.
This equation is the (first) London equation and expresses the acceleration of
electrons under gradients of their ``potential'' $p^\e_0$ and an
electric field. This equation also guarantees that the constraint
(\ref{eq:London0a}) remains automatically satisfied under the
evolution of the electron fluid. 

The two equations (\ref{eq:London0a}) and (\ref{eq:London0b}) were
originally proposed (albeit in their ``orthodox formulation'') by
F.~and H.~London \cite{london35:_em_equ_supercond} and have been very
successful in describing the phenomenology of superconductors, and in
particular the behavior in electric and magnetic fields.

We conclude this section by a ``translation'' into the
orthodox formalism (cf. the discussion of superfluid $^4$He in Paper~I).
Using the entrainment relation (\ref{eq:ElCondEntr}) we can express
the electron momentum $\vp^\e$ as
\begin{equation}
  {\vp^\e\over m_\e} = \vv - {1 \over e n_\S} \vj\,,
\end{equation}
where we have introduced the orthodox \emph{pseudo-density}  $n_\S$ of
superconducting electrons, namely 
\begin{equation}
  n_\S \equiv {n_\e \over 1 - \entrn}\,,\quad \textrm{with} \quad
\entrn \equiv {2 \entr \over m_\e n_\e}\,.
\end{equation}
With this relation, Eq.~(\ref{eq:London0a}) can now be written in its
conventional form as
\begin{equation}
\vB\ = -c \vnabla \times ( \lambda \vj )\,,\quad\textrm{with}\quad
\lambda \equiv {m_\e \over  e^2 n_\S}\,,
\label{eq:LondonII}
\end{equation}
where we have assumed that the background of ions is stationary and
non-rotating, so $\vnabla\times \vv=0$ and $\partial_t
\vv=0$. Therefore  Eq.~(\ref{eq:London0b}) can now be written as
\begin{equation}
  \vE = \partial_t (\Lambda \vj) + {1\over e} \vnabla p^\e_0 \,,
  \label{eq:LondonI}
\end{equation}
where the ``partial pressure'' term $\vnabla p^\e_0=-\vnabla(\mu^\e
-{1\over2}m_\e v_\e^2 + \vv_\e\cdot\vp^\e)$ is often neglected.
Eqs.~(\ref{eq:LondonII}) and (\ref{eq:LondonI}) represent the orthodox
formulation of the classic London equations as usually found in the
superconductivity literature (e.g. see 
\cite{london50:_superfl_supercond,tilley90:_super,tinkham96:_supercond}).

\begin{acknowledgments}
  I would like to thank Brandon Carter and David Langlois for many
  valuable discussions about the relativistic variational principle
  and superfluids. I am also very grateful to Greg Comer and Nils
  Andersson for many helpful comments.

  I acknowledge support from the EU Programme 'Improving the Human 
  Research Potential and the Socio-Economic Knowledge Base' (Research
  Training Network Contract HPRN-CT-2000-00137).
\end{acknowledgments}

\appendix

\section{Approximate Galilean-invariance of electrodynamics}
\label{sec:appr-galil-invar}

As pointed out in the introduction, the combined electro-hydrodynamics
constructed in this paper makes no claim at being strictly
Galilean-invariant. The underlying framework should be thought of as a
fully (locally) Lorentz-invariant theory (as developed in
\cite{carter98:_relat_supercond_superfl,carter02:_super_super_neutr_stars}), 
of which we consider only the small-velocity, low-frequency regime up
to and including effects of order $\O\left(c^{-1}\right)$. It is well-known (e.g. see 
\cite{MTW}) that the first post-Newtonian corrections to uncharged
particle-mechanics in a gravitational field are of order
$\O\left(c^{-2}\right)$, and the corresponding limiting theory is the
strictly Galilean-invariant classical mechanics. This is not the case
for electrodynamics. As pointed out in \cite{bellac73:_galil_elect},
one cannot obtain a Galilean-invariant limit and keep the full
Maxwell-equations, except for introducing the infamous ether.
However, as we will show here, by restricting ourselves to a suitable
``non-relativistic'' regime of small velocities and low frequencies,
the full framework of electrodynamics admits an ``approximately''
Galilean invariant formulation up to and including order $\O(c^{-1})$. 
This will still hold true for the combined
electro-hydrodynamics, as there are no $\O(c^{-1})$ effects entering
from the mechanical sector of the theory. By consistently neglecting 
terms of order $\O(c^{-2})$, the combined theory can be considered 
as ``approximately'' Galilean-invariant in this sense.

The well-known Lagrangian density $\Lagr_\EM$ of the electrodynamic
field $F_{\mu \nu} = 2\nabla_{[\mu}A_{\nu]}$ can be equivalently
expressed in terms of the two (frame-dependent) vectors $\vE$ and
$\vB$, namely
\begin{equation}
  \label{eq:1}
  \Lagr_\EM = {1\over 16 \pi} F_{\mu \nu} F^{\nu \mu} 
= {1\over 8\pi}\left( \vE^2 - \vB^2 \right)\,,
\end{equation}
where $\vE$ and $\vB$, defined for an observer $u^\mu$, are given by
\begin{equation}
  \label{eq:2}
  E^\mu = F^{\mu \nu} u_\nu\,,\quad B^\mu = - {1\over2}
  \epsilon^{\mu\nu\lambda\gamma} F_{\lambda\gamma} u_\nu\,.
\end{equation}
Note that both vectors are purely spatial and orthogonal to $u^\mu$.
In the Lorentz-frame defined by $u^\mu$ they can therefore be
identified with the common 3-vectors used in the ``3+1'' Newtonian
language describing the electric and magnetic field.
Conversely, the field-tensor $F_{\mu\nu}$ is uniquely specified
in terms of $\vE$ and $\vB$, namely 
\begin{equation}
  \label{eq:16}
  \form{F} = \form{u} \wedge \form{E} + ^*\left(\vec{u} \wedge \vB\right)\,,
\end{equation}
where $*$ denotes the Hodge duality operation with respect to the
4-dimensional Levi-Civita tensor $\form{\eps}$, and $\wedge$ stands for the exterior product. 
In the Lorentz-frame $u^\mu$ we can write this in components as
\begin{equation}
  \label{eq:3}
  F_{\mu \nu} = \left( \begin{array}{c c c c}
      0   & -E_x & -E_y & -E_z \\
      E_x & 0    & B_z  & -B_y \\
      E_y & -B_z & 0    &  B_x \\
      E_z &  B_y & -B_x &  0   \\
\end{array}\right)\,.
\end{equation}

Now consider the effect of the transformation to a frame $K'$ moving
with velocity $\vV$ relative to the original frame $K$.
Introducing the symbols $\vb \equiv \vV/c$, $\beta = |\vb|$, $\vn =
\vb/\beta$, and $\gamma = \left( 1 - \beta^2\right)^{-1/2}$,
the corresponding Lorentz-transformation matrix $\Lambda$ can be written as
\begin{eqnarray}
  \label{eq:11}
  {\Lambda^{0'}}_0 &=& \gamma\,, \nonumber\\
  {\Lambda^{0'}}_j &=& {\Lambda^{j'}}_0 = - \beta \gamma \,n^j\,,\\
  {\Lambda^{i'}}_j &=& (\gamma-1) n^j n^k + \delta^{j k}\,.\nonumber
\end{eqnarray}

Using this, we can transform the frame-components of $F_{\mu\nu}$ to
$K'$ and translate them back into $\vE'$ and $\vB'$, which yields (e.g. see
\cite{MTW}) 
\begin{eqnarray}
  \vE'_\parallel &=& \vE_\parallel\,,\quad 
  \vE'_\perp = \gamma\left(\vE_\perp + \vb \times \vB\right)\,,   \label{eq:5a}\\
  \vB'_\parallel &=& \vB_\parallel\,,\quad
  \vB'_\perp = \gamma\left(\vB_\perp - \vb \times \vE \right)\,,  \label{eq:5b}
\end{eqnarray}
where the parallel ($_\parallel$) and orthogonal ($_\perp$)
projections refer to the boost-direction $\beta$.

The range of validity of the formalism developed in this paper is the
``non-relativistic'' small-velocity, low-frequency regime, so we introduce
the small parameter $\ee \ll 1$ and require that all
characteristic velocities $\vec{v}$ satisfy $|\vec{v}|/c \sim
\O(\ee)$, and that time-derivatives of field-quantities are
small compared to spatial derivatives, i.e. $\partial_t B/c \sim 
\partial_t E/c \sim \O(\ee)$. Consequently we need to restrict ourselves to
small Lorentz-boosts, i.e. we assume $\beta \sim \O(\ee)$, and so
\begin{equation}
  \label{eq:6}
  \gamma = \left( 1 - \beta^2 \right)^{-1/2} = 1 + \O(\ee^2)\,.
\end{equation}
We can see that the Lorentz-transformation (\ref{eq:5a},\ref{eq:5b}) of the
electromagnetic fields can now be written as
\begin{eqnarray}
  \vE' &=& \vE + \vb \times \vB + \O(\ee^2)\,,   \label{eq:8a}\\
  \vB' &=& \vB - \vb \times \vE + \O(\ee^2)\,.   \label{eq:8b}
\end{eqnarray}
As pointed out in \cite{bellac73:_galil_elect}, these transformations
do not form a strict Galilean invariance group, because they fail to be
additive. But they do form an ``approximate'' invariance-group up to
order $\O(\ee^2)$ in the sense discussed above. Namely, combining two
boosts, $\vb_1$ and $\vb_2$, we find 
\begin{eqnarray}
  \label{eq:12}
  \vE'' &=& \vE' + \vb_2\times\vB' = \vE + (\vb_1+\vb_2)\times\vB + \O(\ee^2)\,,\\
  \vB'' &=& \vB' - \vb_2\times\vE' = \vB - (\vb_1+\vb_2)\times\vE + \O(\ee^2)\,.
\end{eqnarray}
Let us consider the effect of the transformation
(\ref{eq:8a},\ref{eq:8b}) on the electrodynamics field-Lagrangian (\ref{eq:1}), for
which we easily find 
\begin{equation}
  \label{eq:9}
  \Lagr'_\EM = {1\over8\pi}\left(\vE'^2 - \vB'^2\right) =
  {1\over8\pi}\left(\vE^2-\vB^2\right) + \O(\ee^2)\,.
\end{equation}
In principle this would conclude our demonstration, as both the
Maxwell-equations and the Lorentz-force law are derivable from this
Lagrangian, but for completeness we will also discuss their explicit
transformation properties.

Note that in the conventional Newtonian ``3+1'' language, the boost
$\vV$ results in the following transformations (using the fact that
$x^0=ct$) up to order $\O(\ee^2)$:
\begin{eqnarray}
  \label{eq:10}
  t'   &=&  t\,,\\
  \vx' &=& \vx - \vV\,t\,,\\
  \vv' &=& \vv - \vV\,, \\
  \partial'_t &\equiv& \left.\partial_t\right|_{\vx'} = \partial_t + \vV\cdot\vnabla\,,
\end{eqnarray}
which are just the usual Galilean boost transformations.
Similarly, we obtain the transformation law up to $\O(\ee^2)$ for the
gauge-field vector $A_\mu = (\Ao, \vA)$ as
\begin{eqnarray}
  \label{eq:8}
  \Ao' &=& \Ao + \vb\cdot\vA\,,\\
  \vA' &=& \vA + \Ao \vb\,,
\end{eqnarray}
and one can easily verify using the definition (\ref{eq:DefEMFields}) of
$\vE$ and $\vB$ in terms of $(\Ao, \vA)$ that this is consistent with
the transformations (\ref{eq:8a},\ref{eq:8b}) up to corrections of $\O(\ee^2)$.
It is also interesting to note the transformation properties of the
charge-density $\jo$ and the electrical current $\vj$. For an
individual constituent $\X$, the 4-current $j_\X^\mu = n_\X u_\X^\mu$,
with the 4-velocity $u_\X^\mu=(c, \vv_\X)+\O(\ee^2)$. We therefore
find up to order $\O(\ee^2)$
\begin{equation}
  \label{eq:13}
  \jo_\X' = \jo_\X\,,\quad \vj_\X' = \vj_\X - \jo_\X \vV\,,
\end{equation}
and by the definition (\ref{eq:DefSigJ}) of the total charge-density
$\jo$ and current $\vj$ one easily finds\footnote{This shows that the
  current density \emph{always} transforms according to the ``electric limit''
  as defined in \cite{bellac73:_galil_elect},
  irrespective of the presence of net charges. The authors overlooked
  the fact that even a space-like current has to be the sum of time-like
  ``elementary'' currents, and so the ``magnetic limit'' will never apply
  in real systems.}   
\begin{equation}
  \label{eq:14}
  \jo' = \jo\,,\quad \vj' = \vj - \jo\,\vV\,.
\end{equation}

Applying the transformation (\ref{eq:8a}--\ref{eq:8b}) and
(\ref{eq:10}) to the Maxwell-equations (\ref{eq:Maxwell1}),
(\ref{eq:Maxwell2}), we obtain 
\begin{eqnarray}
  \vnabla\cdot\vB' &=& \vb\cdot(\vnabla\times\vE) = -\vb\cdot (\partial_t \vB /c ) = \O(\ee^2)\,,  \\
  \vnabla\times\vE' + {1\over c}\partial'_t \vB' &=& \vnabla\times\vE
      + {1\over c}\partial_t\vB -\vb\times{1\over c}\partial_t \vE -
      (\vb\cdot\vnabla)(\vb\times\vE) + \O(\ee^2) \nonumber \\ 
      &=& \left[ \vnabla\times\vE + {1\over c}\partial_t\vB \right] + \O(\ee^2)\,,\\
      \vnabla\cdot\vE' - 4\pi \jo' &=& \left[ \vnabla\cdot\vE -
      4\pi\jo\right] + \vnabla\cdot(\vb\times\vB) =
      -\vb\cdot(\vnabla\times\vB) = \O(\ee^2)\,,\\
  \vnabla\times\vB' - {1\over c}\partial'_t\vE' - {4\pi\over c}\vj' &=&
      \vnabla\times\vB - \vb (\vnabla\cdot\vE) - {1\over c}\partial_t \vE
      - \vb \times {1\over c}\partial_t\vB - (\vb\cdot\vnabla)(\vb\times\vB) \nonumber \\
      && - \left( {4\pi\over c}\vj - 4\pi \jo \vb \right) + \O(\ee^2) \nonumber\\
  &=& \left[\vnabla \times \vB - {1\over c}\partial_t \vE - {4\pi\over c}\vj\right] + \O(\ee^2)\,,
\end{eqnarray}
where in the last equation we used the transformation property of the
electrical current density $\vj' = \vj - \jo \vV$.

Finally, the expression for the Lorentz-force (\ref{eq:Fem})
transforms as
\begin{equation}
  \label{eq:7}
  {\vf'_\EM\over n q} = \vE' + {\vv'\over c}\times \vB' = \vE +
  {\vv\over c}\times \vB - {\vv \over c}\times \vb \times \vE +\O(\ee^2) =
  {\vf\over n q} + \O(\ee^2)\,,
\end{equation}
which concludes our demonstration of the approximate Galilean
invariance of the equations.

\bibliography{biblio}

\end{document}